\documentclass[twocolumn,twoside,slac_two]{revtex4}
\usepackage{graphicx}
\usepackage{fancyhdr}
\begin{document}

\title{Experimental Review of Exclusive Semileptonic $B$ Meson Decays and Measurements of $|V_{ub}|$}

%

\author{K.E. Varvell}
\affiliation{School of Physics, The University of Sydney, NSW 2006, Australia}

\begin{abstract}
  We review the current status of experimental measurements of
  the branching fractions for exclusive semileptonic decays of $B$
  mesons to charmless hadrons and the determination of the
  Standard Model CKM parameter $|V_{ub}|$ from these measurements.
\end{abstract}

\maketitle

\thispagestyle{fancy}


\section{Introduction}

The Standard Model (SM) of particle physics contains a number of
parameters whose values are not predicted by theory and must therefore
be measured by experiment. In the quark sector, the elements of the
$3 \times 3$ Cabibbo-Kobayashi-Maskawa (CKM) matrix \cite{ckm} govern
the weak transitions between quark flavours, and precision
measurements of their values are desirable. In particular, much
experimental and theoretical effort is currently being employed to
test the consistency of one of the six ``Unitarity Triangles'' arising
from the CKM formalism, the one most relevant to the decays of $B$
mesons.

Since the first reports of $CP$ violation in the $B$ sector by the
$B$-factory experiments BaBar and Belle in 2001
\cite{babar_cp} \cite{belle_cp}, the precision to which the angle
$\sin 2\phi_1$ ($\sin 2\beta$) characterising indirect $CP$ violation in 
$b \to c \overline{c} s$ transitions has improved to approximately
4\%. This makes a precision measurement of the length of the side of
the Unitarity triangle opposite to $\sin 2\phi_1$ particularly
important as a consistency check of the SM picture. The length of this
side is determined to good approximation by the ratio of the
magnitudes of two CKM matrix elements, $|V_{ub}|/|V_{cb}|$. Both of
these can be measured using exclusive semileptonic $B$ meson decays.
Using charmed semileptonic decays, the precision to which $|V_{cb}|$
has been determined is of order 2\%. On the other hand $|V_{ub}|$,
which can be measured using charmless semileptonic decays, is the most
poorly known of the CKM matrix elements. Both inclusive and exclusive
methods of measuring $|V_{ub}|$ have been pursued, with the inclusive
methods giving a value to something like 7-8\% precision. A review of
the determination of $|V_{ub}|$ using inclusive methods can be found
elsewhere in these proceedings \cite{barberio_talk}. The exclusive
determination of $|V_{ub}|$ currently has a precision poorer than
10\%. The aim of the ongoing programme of measurements reviewed here is to 
improve this precision to better than 5\%, for comparison with the
inclusive results, which have somewhat different experimental and
theoretical systematics, and to provide a sharp consistency test with
the value of $\sin 2\phi_1$.

\section{Exclusive Charmless Semileptonic $B$ Meson Decays}

Measurements of exclusive charmless semileptonic $B$ meson decays, which have
branching fractions of the order $10^{-4}$, can most
readily be made at electron-positron storage rings, where large numbers of
$B \overline{B}$ pairs are produced through the process
$e^+ e^- \to \Upsilon(4S) \to B\overline{B}$. The aim is to measure
the rate of the tree level quark transition $b \to u \ell \nu$, whose
amplitude depends on $V_{ub}$. The situation is complicated by strong
interaction effects, since the $b$ and $u$ quarks are bound into
mesons and form factors depending on $q^2$, the square of the 4-momentum
transferred to the lepton pair, are required. The most promising
decays for measuring $|V_{ub}|$ are those where the final state
meson is spinless, since in this case only two form factors are
required to describe the branching fraction, and one if the mass of
the final state leptons is neglected. For example, if the final state
meson is a pion, the differential branching fraction can be written
in the form
\begin{equation}
  \frac{d\Gamma \left(B \to \pi \ell \nu \right)}{dq^2}
  = \frac{G_F^2}{24\pi^3} |V_{ub}|^2 p_\pi^3 |f_+( q^2 ) |^2
\label{eq-pibf}
\end{equation}
where $p_\pi$ is the pion momentum and $f_+(q^2)$ the form factor.
Thus experiment determines the product $|V_{ub}| |f_+(q^2)|$, and to
extract $|V_{ub}|$, both the shape and normalization of $f_+(q^2)$
are required. Input on $f_+(q^2)$ has come from
theory, firstly in the form of quark-model predictions \cite{isgw2}.
More recently lattice QCD calculations such as
those of the HPQCD \cite{hpqcd-06} and FNAL \cite{fnal-04}
Collaborations, and calculations based on Light Cone Sum Rules
(LCSR) \cite{ball-and-zwicky-05} have become available. These latter
two approaches are complementary in the sense that lattice predictions
are applicable at high $q^2 > 16\ \mathrm{GeV}^2/c^2$, whereas LCSR
predictions are applicable at lower $q^2 < 14\ \mathrm{GeV}^2/c^2$.
The lattice predictions are now based on unquenched calculations.
Experiments have traditionally employed a parametrization of the form
factor shapes to extend the predictions to the full $q^2$ range for
which they have data.

More discussion on the issue of the calculation of form factors, the
need or otherwise for extrapolation and the effect of form factors on
the determination of $|V_{ub}|$ can be found in
\cite{hill-05} \cite{mackenzie_talk}.
Since experiments are reaching the point where they
can begin to provide information on the form factor shapes, the
dominant issue in extracting $|V_{ub}|$ is the form factor normalization.

\section{Experimental Methods for Determining Branching Fractions}

All recent results on exclusive charmless semileptonic $B$ meson
decays have come from the $e^+ e^-$ machines CESR at Cornell, PEP-II at
SLAC and KEKB at KEK, and their respective experiments CLEO, BaBar and
Belle. In each case, the required decay must be identified in final
states which also contain the decay products of the other $B$ meson, and
without direct detection of the neutrino. Suppression of combinatorial
backgrounds, backgrounds from the $b \to c \ell \nu$ process, and
backgrounds from the underlying continuum to the $\Upsilon(4S)$
resonance is the major challenge. Three methods of signal extraction
have been used in the measurements tabulated here, which we now
briefly describe.

\subsection{Untagged Method}

The pioneering CLEO measurements developed the untagged method which
was the only method available given the size of their dataset and the
rareness of the decays under study. Since the total 4-momentum of the
final state is fixed by the decaying $\Upsilon(4S)$, the 4-momentum of
the escaping neutrino is estimated by subtracting the visible
4-momentum from the known total, attributing this missing 4-momentum to a
neutrino. Candidate $B \to X_u \ell \nu$ decays are then selected on
the basis of two variables, the beam constrained mass $m_{BC}$ and
$\Delta E$, defined as follows
\begin{eqnarray}
  m_{BC}   & = & \sqrt{E_{\mathrm{beam}}^2 - p_B^2} \\
  \Delta E & = & E_B - E_{\mathrm{beam}}
\label{mbc_de}
\end{eqnarray}
Here $E_{\mathrm{beam}}$ is the beam energy in the center of mass frame
  of the $\Upsilon(4S)$, and $E_B$ and $p_B$ are the energy and
  magnitude of momentum of the $B$ candidate in the same frame.
  For $B \to X_u \ell \nu$ decays these will ideally have values close to
  the $B$ meson mass and zero respectively, whilst for background
  this is not the case. The method is denoted untagged since the
  $B$ meson recoiling against the signal candidate $B$ is not
  explicitly reconstructed.

The major advantage of this method over the two to be described below
is a relatively high efficiency (of the order of several percent).
The major disadvantage is that the resolution of the neutrino
4-momentum is relatively poor, and this results in a lower
purity and signal to background ratio.

\subsection{Semileptonic Tagging}

Semileptonic tagging involves the partial reconstruction of a
semileptonic $B$ meson decay to charm recoiling against the
signal $B \to X_u \ell \nu$  candidate. Several $D$ and $D^{*}$ decay
modes are used for the tagging. Since the final state contains
a neutrino from both signal and tagging $B$, kinematic constraints
must be employed to separate signal events from background.
Backgrounds are lower than for the untagged method,
but so is efficiency.

\subsection{Full Reconstruction Tagging}

In this method the $B$ meson recoiling against the signal $B$
candidate is fully reconstructed in a selected set of hadronic $B$ decay modes
containing a charmed meson. A sizeable number of modes are used for the
tagging. In this case neutrino 4-momentum resolution is excellent and
very low backgrounds result. The major disadvantage of the method over
the two just described is very low tagging efficiency (typically a
fraction of a percent). 

The $B$-factory experiments Belle and BaBar have collected of
order $500 \times 10^6$ and $300 \times 10^6\ B\overline{B}$ pairs
respectively to date. These represent very large datasets, which will
continue to grow in the next few years. This will offset the
disadvantage of low efficiency for the full reconstruction tagging
method, and it will become the method of choice.

\section{$\mathbf{B \to \pi \ell \nu}$ Branching Fractions}

\begin{figure}[h]
\centering
\includegraphics[width=80mm]{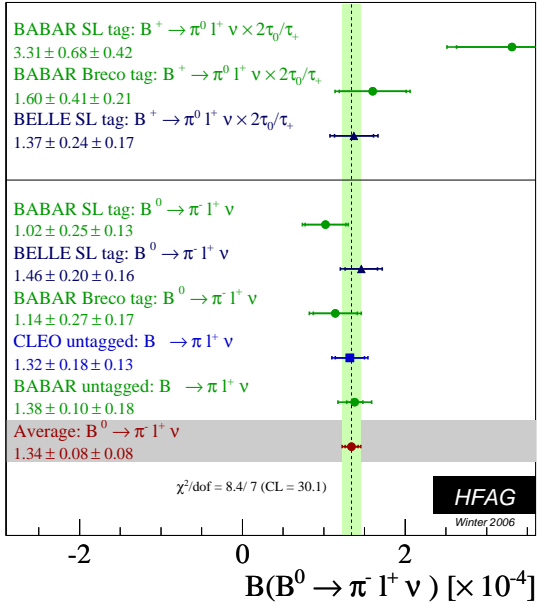}
\caption{Branching fraction measurements for exclusive
  $B \to \pi \ell \nu$ decays, reproduced from the HFAG compilation
\protect\cite{hfag-winter-06} and discussed in this review.}
\label{pilnu_hfag}
\end{figure}

Table~\ref{b_to_pi} lists the current measurements of the branching
fraction for $B \to \pi \ell \nu$, for both charged and neutral pion modes.
These are displayed in Figure~\ref{pilnu_hfag}, which is reproduced
from the HFAG Winter 2006 compilation \cite{hfag-winter-06}. In this
figure, results for the $\pi^0 \ell \nu$ mode have been multiplied by
a factor of two to reflect isospin expectations, and corrected for the
difference in lifetimes of charged and neutral $B$ mesons. Note that
presently at least, the untagged methods still give the best
experimental precision for the branching fraction.

%

\begin{table}[h]
\begin{center}
  \caption{Measurements of branching fractions of exclusive
    $B \to \pi \ell \nu$ decay modes. In each case, the first error is
    statistical, the second experimental systematic, the third due to
    form factor uncertainties for the signal mode, and the fourth, when
    present, due to form factor uncertainties for crossfeed modes.
    U indicates untagged method, S semileptonic tagging method, and F
    full reconstruction tagging. 
    The result in the last row of the table
    combines the previous two, using isospin relations.}
\begin{tabular}{|l|c|c|c|}
  \hline \textbf{Expt/Tag} & \textbf{Mode} &
  \textbf{$B\overline{B}$} &
\textbf{Branching Fraction} \\
& &  
\textbf{$\left[ 10^{6} \right]$} &
\textbf{$\left[ 10^{-4} \right]$}
\\
\hline
CLEO \cite{cleo-untagged-03} \hfill U &  $B^0 \to \pi^- \ell \nu$ &
9.7 & 
$1.33 \pm 0.18 \pm 0.11
\pm 0.01$ \\
& & & $\pm 0.07$ \\
\hline
BaBar \cite{babar-untagged-05} \hfill U & $B^0 \to \pi^- \ell \nu$ &
86 &
$1.38 \pm 0.10 \pm 0.16
      \pm 0.08$ \\
\hline
Belle \cite{belle-sl-06} \hfill S & $B^0 \to \pi^- \ell \nu$ &
275 & 
$1.38 \pm 0.19 \pm 0.14
      \pm 0.03$ \\
\hline
Belle \cite{belle-sl-06} \hfill S  & $B^+ \to \pi^0 \ell \nu$ &
275 & 
$0.77 \pm 0.14 \pm 0.08
      \pm 0.00$ \\
\hline
BaBar \cite{babar-sl-pip-05} \hfill S  & $B^0 \to \pi^- \ell \nu$ &
232 & 
$1.03 \pm 0.25 \pm 0.13$ \\
\hline
BaBar \cite{babar-sl-piz-05} \hfill S  & $B^+ \to \pi^0 \ell \nu$ &
88 & 
$1.80 \pm 0.37 \pm 0.23$ \\
\hline
BaBar \cite{babar-fullrecon-pi-05} \hfill F & $B^0 \to \pi^- \ell \nu$ &
233 & 
$1.14 \pm 0.27 \pm 0.17$ \\
\hline
BaBar \cite{babar-fullrecon-pi-05} \hfill F & $B^+ \to \pi^0 \ell \nu$ &
233 & 
$0.86 \pm 0.22 \pm 0.11$ \\
\hline
BaBar \cite{babar-fullrecon-pi-05} \hfill F & $B \to \pi \ell \nu$ &
233 & 
$1.28 \pm 0.23 \pm 0.16$ \\
\hline
\end{tabular}
\label{b_to_pi}
\end{center}
\end{table}

The $B \to \pi \ell \nu$ branching fraction averaged over all of
these measurements is quoted by HFAG to be
$\left[1.34 \pm 0.08(\textit{stat}) \pm 0.08(\textit{syst})
\right] \times 10^{-4}$, which represents an experimental precision of
around 8\%.

\begin{figure}
\includegraphics[width=65mm]{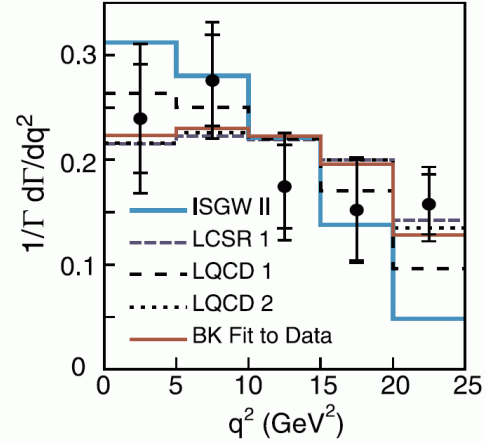}%
\caption{{Shape dependence of the partial branching fraction for
  $B^0 \to \pi^- \ell \nu$ obtained by BaBar using their untagged analysis
\protect\cite{babar-untagged-05}.}}
\label{babar_untagged_pip_q2}
\end{figure}

\begin{figure}
\includegraphics[width=60mm]{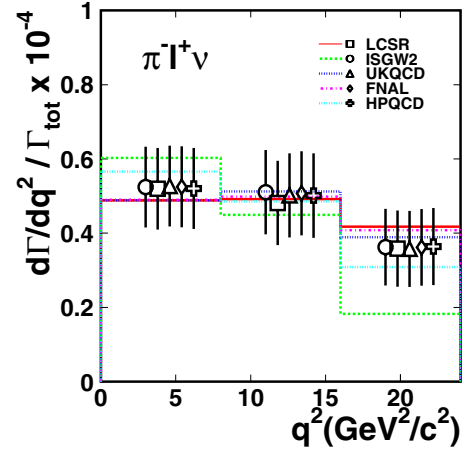}%
\caption{Shape dependence of the partial branching fraction for
  $B^0 \to \pi^- \ell \nu$ obtained by Belle using their semileptonic
  tag analysis \protect\cite{belle-sl-06}.}
\label{belle_sl_pip_q2}
\end{figure}

Most of the measurements of $\mathcal{B}(B \to \pi \ell \nu)$
now provide some information on the $q^2$ dependence. By way of
illustration, Figure~\ref{babar_untagged_pip_q2} shows the shape
dependence of the partial branching fraction for $B^0 \to \pi^- \ell \nu$
obtained by BaBar in 5 $q^2$ bins using their untagged analysis
\cite{babar-untagged-05} based on $86 \times 10^6 B\overline{B}$ pairs,
whilst Figure~\ref{belle_sl_pip_q2} reproduces the same quantity
obtained by Belle in 3 $q^2$ bins using their semileptonic tag
analysis \cite{belle-sl-06} based on $275 \times 10^6 B\overline{B}$
pairs. In each case predictions based on different assumptions
about the signal form factor dependence are shown. The
BaBar data appears to be able to rule out a form factor shape given
by simple quark model calculations. The Belle data also shows the
effect of the form factor model on the extracted value of the
branching fractions, which enters primarily through the Monte Carlo
estimation of the detection efficiency.

\section{$|V_{ub}|$ from $\mathbf{B \to \pi \ell \nu}$ Decays}

\begin{table*}[t]
\begin{center}
  \caption{Measurements of branching fractions of exclusive
    $B \to X_u \ell \nu$ decay modes other than $B \to \pi \ell \nu$.
    In each case, the first error is
    statistical, the second experimental systematic, the third due to
    form factor uncertainties for the signal mode, and the fourth, when
    present, due to form factor uncertainties for crossfeed modes.
    U indicates untagged method, S semileptonic tagging method, and F
    full reconstruction tagging.}
\begin{tabular}{|l|c|c|c|}
  \hline \textbf{Expt/Tag} & \textbf{Mode} &
  \textbf{$B\overline{B}$} &
\textbf{Branching Fraction} \\
& &  
\textbf{$\left[ 10^{6} \right]$} &
\textbf{$\left[ 10^{-4} \right]$}
\\
\hline
CLEO \cite{cleo-untagged-03} \hfill U &  $B^0 \to \rho^- \ell \nu$ &
9.7 &
$2.17 \pm 0.34 ^{+0.47}_{-0.54}
\pm 0.41 \pm 0.01$ \\
\hline
CLEO \cite{cleo-untagged-00} \hfill U &  $B^0 \to \rho^- \ell \nu$ &
3.3 &
$2.69 \pm 0.41 ^{+0.35}_{-0.40}
\pm 0.50$ \\
\hline
BaBar \cite{babar-fullrecon-04} \hfill F & $B^0 \to \rho^- \ell \nu$ &
88 &
$2.57 \pm 0.52 \pm 0.59$ \\
\hline
BaBar \cite{babar-untagged-03} \hfill U & $B^0 \to \rho^- e \nu$ &
55 &
$3.29 \pm 0.42 \pm 0.47
\pm 0.60$ \\
\hline
BaBar \cite{babar-untagged-05} \hfill U & $B^0 \to \rho^- \ell \nu$ &
83 &
$2.14 \pm 0.21 \pm 0.51
\pm 0.28$ \\
\hline
Belle \cite{belle-sl-06} \hfill S & $B^0 \to \rho^- \ell \nu$ &
275 &
$2.17 \pm 0.54 \pm 0.31
\pm 0.08$ \\
\hline
CLEO \cite{cleo-untagged-03} \hfill U &  $B^+ \to \eta \ell \nu$ &
9.7 &
$0.84 \pm 0.31 \pm 0.16
\pm 0.09$ \\
\hline
Belle \cite{belle-sl-06} \hfill S & $B^+ \to \rho^0 \ell \nu$ &
275 & 
$1.33 \pm 0.23 \pm 0.17
\pm 0.05$ \\
\hline
Belle \cite{belle-untagged-omega-04} \hfill U & $B^+ \to \omega \ell \nu$ &
85 &
$1.3 \pm 0.4 \pm 0.2
\pm 0.3$ \\
\hline
\end{tabular}
\label{b_to_other}
\end{center}
\end{table*}

Given a branching fraction measurement, $|V_{ub}|$ can be estimated
using the relation
\begin{equation}
  |V_{ub}| = \sqrt{\frac{\mathcal{B}\left( B \to \pi \ell^+ \nu
      \right)}{{\tilde{\Gamma}}_{thy} \tau_B}}
\label{vub}
\end{equation}
where ${\tilde{\Gamma}}_{thy}$ is the form factor normalization
provided by theory and $\tau_B$ is the $B$ proper lifetime.

Most of the analyses listed in Table~\ref{b_to_pi} have produced values for
$|V_{ub}|$, based either on partial branching fraction values over
the limited $q^2$ region for which selected theoretical predictions
for the form factor normalizations are applicable, or by using the
full $q^2$ region and an extrapolation of the predictions to the full
region. Some analyses quote values for both scenarios. In some
cases the charged and neutral pion channels are combined using isospin
relations to improve the statistical precision. In general the
precision of the extracted $|V_{ub}|$ values is dominated by the form
factor uncertainties, with the overall precision for $|V_{ub}|$
tending to be better when a limited $q^2$ range is employed. 

The individual $|V_{ub}|$ values are not tabulated here. Details
can be found in the referenced papers. The situation can be summarised
by quoting the approach taken by the HFAG.
The branching fraction measurements from all of the analyses are
combined to give a global average in three $q^2$ ranges --- the full
range which extends to approximately $25\ \mathrm{GeV}^2/c^2$, and the
ranges $q^2 < 16\ \mathrm{GeV}^2/c^2$, to which LCSR apply, and
$q^2 > 16\ \mathrm{GeV}^2/c^2$, to which Lattice QCD applies.
$|V_{ub}|$ is then calculated using the predictions of LCSR
\cite{ball-and-zwicky-05}, HPQCD lattice \cite{hpqcd-06},
FNAL lattice \cite{fnal-04} and their stated theoretical errors.

At this point in time a global average based on exclusive decays is
not quoted by the HFAG \cite{hfag-winter-06}. They obtain the
following values in the partial $q^2$ ranges
\begin{eqnarray*}
  |V_{ub}| & = & 3.25 \pm 0.17 ^{+0.54}_{-0.36} \quad
  \mathrm{LCSR\ \ }\ q^2 < 16\ \mathrm{GeV}^2/c^2 \\
  |V_{ub}| & = & 4.44 \pm 0.30 ^{+0.67}_{-0.46} \quad
  \mathrm{HPQCD }\ q^2 > 16\ \mathrm{GeV}^2/c^2 \\
  |V_{ub}| & = & 3.76 \pm 0.25 ^{+0.65}_{-0.43} \quad
  \mathrm{FNAL\ \ }\ q^2 > 16\ \mathrm{GeV}^2/c^2
\end{eqnarray*}
which illustrate the need to reduce the size of the theory errors,
quoted second. The first error is a combination of the experimental
statistical and systematic errors. The full methodology is described
in the HFAG paper. When the full $q^2$ range is used, the gain from
reducing the experimental error tends to be offset by a larger
increase in the theory error caused by the extrapolation.

\section{Other Exclusive Charmless Semileptonic Meson Decays}

Whilst $B \to \pi \ell \nu$ decays have been of most recent interest
for the extraction of $|V_{ub}|$, measurements exist for the branching
fraction of other charmless semileptonic decays. These are listed in
Table~\ref{b_to_other}. In many cases these measurements were
performed in conjunction with those for $B \to \pi \ell \nu$, since
this approach allows for the most consistent treatment of crossfeed
between channels.

%




\section{Exclusive Charmed Semileptonic Meson Decays}

Alongside improvements in the determination of $|V_{ub}|$ from
$B \to X_u \ell \nu$ decays, better knowledge of $B \to X_c \ell \nu$
decays is important for two reasons. Further improvement in the
precision of $|V_{cb}|$ helps in constraining the Unitarity Triangle.
Secondly, better knowledge of the branching fractions for these decays
as a function of $q^2$ leads to better knowledge of the shapes of
the relevant form factors, which when used as input to Monte Carlo
simulations of charm backgrounds in $B \to X_u \ell \nu$ measurements
will improve systematic errors.

The BaBar Collaboration have recently released an update to their
determination of $|V_{cb}|$ from $B \to D^* \ell \nu$ decays
\cite{babar-dstar-06}, based on $86 \times 10^6\ B\overline{B}$ pairs.
New values for the parameters $\rho^2$, $R_1$ and $R_2$ characterising
the helicity structure of the decays leads to an improvement of order
25\% in the systematic error on $|V_{cb}|$, which has the new value
$|V_{cb}| = \left[ 37.6 \pm 0.3(\textit{stat}) \pm 1.3(\textit{syst})
^{+1.5}_{-1.3}(\textit{theor}) \right] \times 10^{-3}$.

\section{Summary}

The CLEO, BaBar and Belle experiments have between them now provided
measurements of the semileptonic branching fractions of $B$ mesons to
the following final state mesons: $\pi^\pm$, $\pi^0$, $\rho^\pm$,
$\rho^0$, $\omega$ and $\eta$, employing three methods of identifying
signal decays in the presence of a missing neutrino. The most precise
values for $|V_{ub}|$ come from the charged pion channel, with an
experimental error approaching 5\% precision but with a theory error
arising from imprecise knowledge of the form factor normalization
dominating the overall precision. Information on the shape of the form
factors is now beginning to be provided by experiment. Further
increases in the size of the Belle and BaBar datasets will
allow full reconstruction tagging techniques to be competitive.
Coupled with improvements in lattice calculations, this should in future
see the overall precision on $|V_{ub}|$ approach the desired 5\%
level.

\bigskip 

\end{document}